\documentstyle{mn} 
\input epsf

\title[Galactic wakes]
{Are ``Bondi-Hoyle Wakes'' detectable in clusters of galaxies?}
\author[Irini Sakelliou]
          {Irini~Sakelliou\\ 
Mullard Space Science Laboratory, University College
London, Holmbury St Mary, Dorking, Surrey RH5 6NT, UK\\}

\begin{document}
\maketitle
 
\begin{abstract} 

In clusters of galaxies, the reaction of the intracluster medium (ICM) to the
motion of the co-existing galaxies in the cluster triggers the formation of
unique features, which trace their position and motion. Galactic
wakes, for example, are an apparent result of the ICM/galaxy interactions, and
they constitute an important tool for deciphering the motion of the cluster galaxies.

In this paper we investigate whether Bondi-Hoyle accretion can 
create galactic wakes by focusing the ICM behind moving
galaxies.  The solution of the equations that describe this physical problem
provide us with observable quantities along the wake at any time of its
lifetime.  We also investigate which are the best environmental conditions for
the detectability of such structures in the X-ray images of clusters of
galaxies.

We find that significant Bondi-Hoyle wakes can only be formed in low
temperature clusters, and that they are more pronounced behind slow-moving,
relatively massive galaxies. The scale length of these elongated structures is
not very large: in the most favourable conditions a Bondi-Hoyle wake in a
cluster at the redshift of z=0.05 is 12~arcsec long. However, the wake's X-ray
emission is noticeably strong: the X-ray flux can reach $\sim$30 times the flux
of the surrounding medium. Such features will be easily detectable in {\it
Chandra's} and {\it XMM-Newton's} x-ray images of nearby, relatively poor
clusters of galaxies.

\end{abstract}
 
\begin{keywords}
galaxies: clusters: general -- galaxies: interaction -- intergalactic medium --
galaxies: kinematics and dynamics -- X-rays: galaxies
\end{keywords}

\section{Introduction}

In clusters of galaxies, the interactions of the intracluster medium (ICM) with
a moving cluster galaxy is expected to modify both the local properties of the
surrounding medium and the galaxy itself. One of the manifestations of such
interactions is the Bondi-Hoyle (B-H) accretion (Bondi \& Hoyle 1944). This
physical process can be pictured as follows: as the galaxy is moving in the
cluster, ICM particles are deflected by the galaxy's gravity and concentrate
behind it, into the galactic wake.

Intuitively, one might think that B-H accretion creates overdense and
cool regions of enhanced x-ray emission behind the galaxies. As
a consequence, the hot interstellar media (ISM) of these galaxies 
look as if they have been disfigured: instead of being azimuthally
symmetric they appear elongated, or as if they have a `plume' of
x-ray emission attached to them.  Such elongated features have now been
identified in the x-ray images of clusters of galaxies (e.g., around
NGC~1404 in the Fornax cluster of galaxies; Jones et al. 1997). An
up-to-date list of wake candidates can be found in Stevens, Acreman \&
Ponman (1999).

Bondi-Hoyle accretion is not the only manifestation of the ISM/ICM interactions
which creates elongated features behind moving galaxies.  Ram pressure
stripping can also shape the ISM in such a way that it appears elongated. The
difference in the nature of the two elongated structures is apparent: a B-H
wake comprises of ICM material, while a ram pressure-induced wake contains
galactic material (ISM).

Unfortunately, instrumentation prior to {\it Chandra} and {\it XMM-Newton} has
not generally allowed the separation of the pure galactic and the wake
components, either spatially or spectroscopically. Only the elliptical galaxy
M86 has offered us the opportunity to gain more insight into the nature of its
wake (Rangarajan 1995).  The observational fact that the metallicity of M86's
wake is higher than the metallicity in the surrounding medium, has been used to
assign it a galactic origin. It seems most probable that in the case of M86,
stripping of its ISM is currently in action, and that its wake consists mostly
of galactic material. No other well studied example is currently known. The
danger arising from the inability to separate the wake's emission from the
galaxy itself, is that the analysis of the galaxy's X-ray data would lead to
false conclusions for the characteristics of its X-ray halo.  The dilution of
ISM gas by the ICM gas in the B-H wake could, for example, lead to confusing
results for the metal abundances in elliptical galaxies.

It should be understood that under certain conditions, B-H accretion and ram
pressure stripping may occur simultaneously, creating wakes which contain a
mixture of both galactic and intracluster material. Recent numerical
simulations by Stevens et al.~(1999) have shown that this picture can be indeed
correct. However, it is not clear yet which are the effects of each separate
process, and how the dominance of one process over the other depends
on the environmental parameters. It is expected, though, that dense ICMs and
high galactic speeds favour ram pressure stripping. However, the results of the
B-H accretion cannot be foreseen so easily. If we want to understand the action
of B-H accretion and be able to find the environmental dependences, we have to
study this process in its first principles.

The aim of this paper is to disentangle the two physical processes by studying
the action and results of the B-H accretion. We address questions such as:
under which conditions B-H accretion occurs in clusters of galaxies;
which clusters are the best candidates for detecting B-H wakes; and whether
x-ray observations with the {\it Chandra} and {\it XMM-Newton} observatories
can reveal B-H wakes.

The remainder of this paper is organized as follows: in \S\ref{methodology} we
present the methodology we followed to calculate the properties of B-H
wakes, section \S\ref{simulations} discusses the constraints and input
parameters imposed by the problem itself. The results of the simulations are
presented in \S\ref{results}. Finally, in \S\ref{discussion} our results are
compared to available observations, and simulations of similar physical
processes.

\section{Creation and evolution of a B-H wake}\label{methodology}

Consider a test volume behind a moving galaxy.  As the
galaxy travels in the cluster at the speed of $v_{\rm gal}$, particles
of the ICM are deflected by the galactic gravitational potential, and
directed into this volume. The ICM particles that are influenced
by the galaxy's attraction, and modify their direction of motion, are the
ones contained within a cylinder of radius equal to the accretion
radius ($R_{\rm acc}$) : 
\begin{equation} 
R_{\rm acc}=\frac{2GM_{\rm gal}}{v_{\rm gal}^{2} + c_{s}^{2}} 
\label{acc_radius}
\end{equation}
(Bondi 1952), where $M_{\rm gal}$ and $c_{s}$ are the mass of the galaxy and
the local speed of sound respectively.

\begin{figure}
\begin{center}
 \leavevmode
 \epsfxsize 0.95\hsize 
 \epsffile{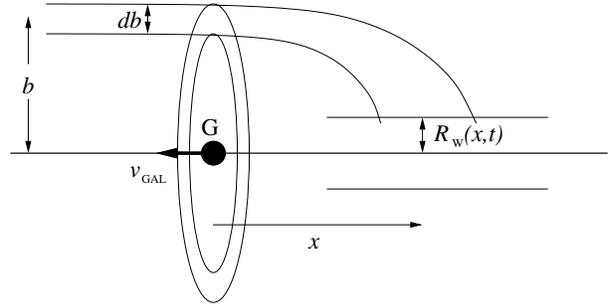} \end{center}
\caption{Schematic diagram of the configuration used for the
calculations of the B-H accretion. The galaxy `G' is moving in
a cluster of galaxies through an ICM of number density $n_{\rm ICM}$.}
\label{schema} 
\end{figure}

In time $dt$, the particles that enter the wake, at a position $x$
along the accretion axis, are the ICM's particles which were initially
in a shell of width $db$ and length $ds=v_{\rm gal} dt$, and had
impact parameter $b$ (see Fig~\ref{schema}):
\begin{equation}
d^{2}z_{\rm acc}(x)=2 \pi b v_{\rm gal} n_{\rm ICM} db \ dt,
\label{z_acc}
\end{equation}
where $n_{\rm ICM}$ is the number density of the surrounding medium.

The particles that start with an impact parameter $b$ land in the wake
along the accretion axis at a distance $x$ from the galaxy. Assuming that the
ballistic approximation is valid (see \S3.1.2) we find that 
the position $x$ is given by:
\begin{equation}
x=\frac{-GM_{\rm gal} + (G^{2} M_{\rm gal}^{2} + v_{\rm
gal}^4 b^{2})^{1/2} }{v_{\rm gal}^{2}}.
\end{equation}
The velocity, $v_{\rm in}(x)$, at the position $x$ on the accretion axis
relates to the initial velocity of the ICM's particles ($v_{\rm gal}$) by:
\begin{equation}
v_{\rm in}(x)=v_{\rm gal} \frac{b}{x}.
\end{equation}

The effect of these incoming ICM's particles is i) the increase
of the internal energy [$E_{\rm int}(x,t)$] of the wake, and ii) its
confinement by the pressure [$P_{\rm acc}(x,t)$]:
\begin{equation}
P_{\rm acc}(x,t) \ dx dt=\frac{\mu m_{\rm p} v_{\rm
in}^{\prime} (x,t)}{2 \pi R_{\rm w}(x,t)} \ d^{2} z_{\rm acc}(x), 
\label{P_acc}
\end{equation}
where $d^{2} z_{\rm acc}(x)$ is given by eq.~\ref{z_acc}, and $R_{\rm
w}(x,t)$ is the radius of the wake at the position $x$
(Fig.~\ref{schema}). In
eq.~\ref{P_acc}, $v_{\rm in}^{\prime}(x,t)$ is the
velocity of the incoming particles in the wake's frame of
reference, and it is given by:
\begin{equation}
v_{\rm in}^{\prime}(x,t) = v_{\rm in}(x) + \frac{dR_{\rm w}(x,t)}{dt}
\end{equation}

As the galaxy travels through the ICM, the constant replenishment of
the wake with particles and energy causes a continuous change to its
properties.  The evolution of the wake is governed by two equations:

\begin{itemize}
\item{the conservation of energy:
\begin{eqnarray}
\lefteqn{\frac{d E_{\rm int}(x,t)}{dt} =}&& \nonumber \\
&&=-L_{\rm bol}(x,t) -P_{\rm w}(x,t)  
\frac{dV_{\rm w}(x,t)}{dt} + \frac{dE_{\rm incom}(x,t)}{dt}
\label{energy}
\end{eqnarray}}
\item{and the conservation of the momentum ($p$): 
\begin{equation}
\frac{dp}{dt} = \int [P_{\rm w}(x,t)-P_{\rm ICM} - P_{\rm acc}(x,t)] d\sigma
\label{momentum}
\end{equation}
where $\sigma$ is the surface containing the wake.}
\end{itemize}

In the above two equations, $P_{\rm w}(x,t)$, and $P_{\rm
ICM}$ are the pressures of the wake and the surrounding ICM
respectively.
In eq.~\ref{energy}, the first term [$L_{\rm bol}(x,t)$] represents the
energy radiated away via thermal bremsstrahlung, and it is calculated in
the appendix~\ref{Lbol}. The
second term is the work done by the wake to the surrounding
medium. The third term is the energy added to the wake by the
accreted particles :
\begin{equation}
dE_{\rm incom}(x,t) = \frac{3}{2} d^{2}z_{\rm acc}(x,t)
kT_{\rm w}(x,t),
\end{equation}
where it was assumed that the incoming particles have enough time to thermalize
with the existing particles in the wake, and reach a Maxwellian distribution
at the temperature of $kT_{\rm w}(x,t)$ (see \S\ref{simulations}).

\section{Simulations}\label{simulations}

Equations \ref{momentum} and \ref{energy} were solved numerically to predict
the temperature [$kT_{\rm w}(x,t)$], and the number density [$n_{\rm w}(x,t)$]
of the wake at any position $x$ along the accretion axis. The time step of the
integration process was constant and such as to allow the gas in the wake to
reach a Maxwellian equilibrium. Knowing the temperature and density at any time
$t$, and position $x$, the luminosity [$L_{\rm bol}(x,t)$] can be found from
equation \ref{brems}.  The central surface brightness along the accretion axis
[$\Sigma_{\rm E_1 - E_2}(0)$], and the wake's surface brightness distribution
are calculated using eq.~\ref{central_surf_bri} and \ref{surf_bri_distr}
respectively.

The initial conditions for the simulations were chosen to comply with the
conditions associated with this problem (Table~\ref{tab:simul}) and are derived
in the next sections.  The chosen values for the parameters of the ICM and the
galaxies' velocity represent the conditions found in poor and moderately rich
clusters of galaxies, which as will become apparent in the next sections, are
the fertile environments for B-H wake creation.

\subsection{Input constraints and initial conditions}

\subsubsection{The mass of the galaxy}

The accretion radius (eq.~\ref{acc_radius}), defines the size of the region
around the cluster galaxy in which the galactic gravitational potential is
strong enough to change the direction of motion of all the ICM particles which
happen to be inside that region, and direct them into the wake. Particles
approaching the galaxy with impact parameters less than $R_{\rm acc}$ are
deflected into the wake, while particles with $b > R_{\rm acc}$ continue their
motion passed the galaxy unaffected. Therefore, an apparent, and simple
condition for a galaxy to have a B-H wake, is that its accretion radius should
be larger than its size.

As is obvious from eq.~\ref{acc_radius}, the potential of a wake being
formed behind a cluster galaxy depends on the velocity and the mass of
the galaxy.  If we assume an average velocity for the galaxies
in a cluster of $v_{\rm gal}= \sqrt{3} \sigma$, where the velocity
dispersion $\sigma$ is given by the $\sigma - T$ relation in clusters
of galaxies: 
\begin{equation} 
\sigma \sim 400 \left( \frac{kT_{\rm
ICM}}{\rm keV} \right)^{0.5}, \; ({\rm km \ s^{-1}}) \label{sigma_T}
\end{equation} 
(White et al. 1997; Wu, Fang \& Xu 1998), and a sound
speed $c_{s}$ of: 
\begin{equation} 
c_{s}=516 \left( \frac{kT_{\rm ICM}}{\rm
keV} \right)^{0.5}, \; ({\rm km \ s^{-1}}) 
\label{sound_v}
\end{equation} 
we find that the condition $R_{\rm acc} \geq R_{\rm
gal}$ gives: 
\begin{equation} \left( \frac{M_{\rm gal}}{\rm gr}
\right) \geq 5.6 \times 10^{22} \left( \frac{kT_{\rm ICM}}{\rm keV}
\right) \left( \frac{R_{\rm gal}}{\rm cm} \right) 
\label{mass_con}
\end{equation}

\begin{figure}
 \begin{center}
 \leavevmode
 \epsfxsize 0.95\hsize
 \epsffile{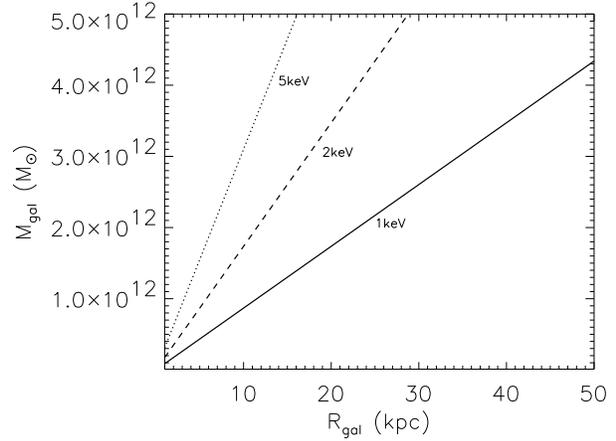}
 \end{center}
\caption{Galaxy mass-radius constraints for B-H accretion in different
temperature environments. The lines are for $R_{\rm acc} = R_{\rm gal}$ for 1
(solid line), 2 (dashed line), and 5~keV (dotted line) temperature
clusters. Equation~\ref{mass_con} is only valid for galaxies which lie in the
space between the lines and the $y$-axis.}
\label{fig:constraints}
\end{figure} 

In Fig.~\ref{fig:constraints} we plot the permitted ranges of galactic masses
and radii for a range of temperatures of the ICM. Galaxies which lie in the
range between the plotted lines and the $y$-axis will produce large-scale B-H
wakes. Figure~\ref{fig:constraints} demonstrates that in high temperature
environments, only very compact objects can have galactic wakes produced by the
B-H accretion. On the other hand, cool environments are more likely to host
galactic wakes.

\begin{figure*}
\begin{center}
 \leavevmode
 \epsfxsize 0.47\hsize 
 \epsffile{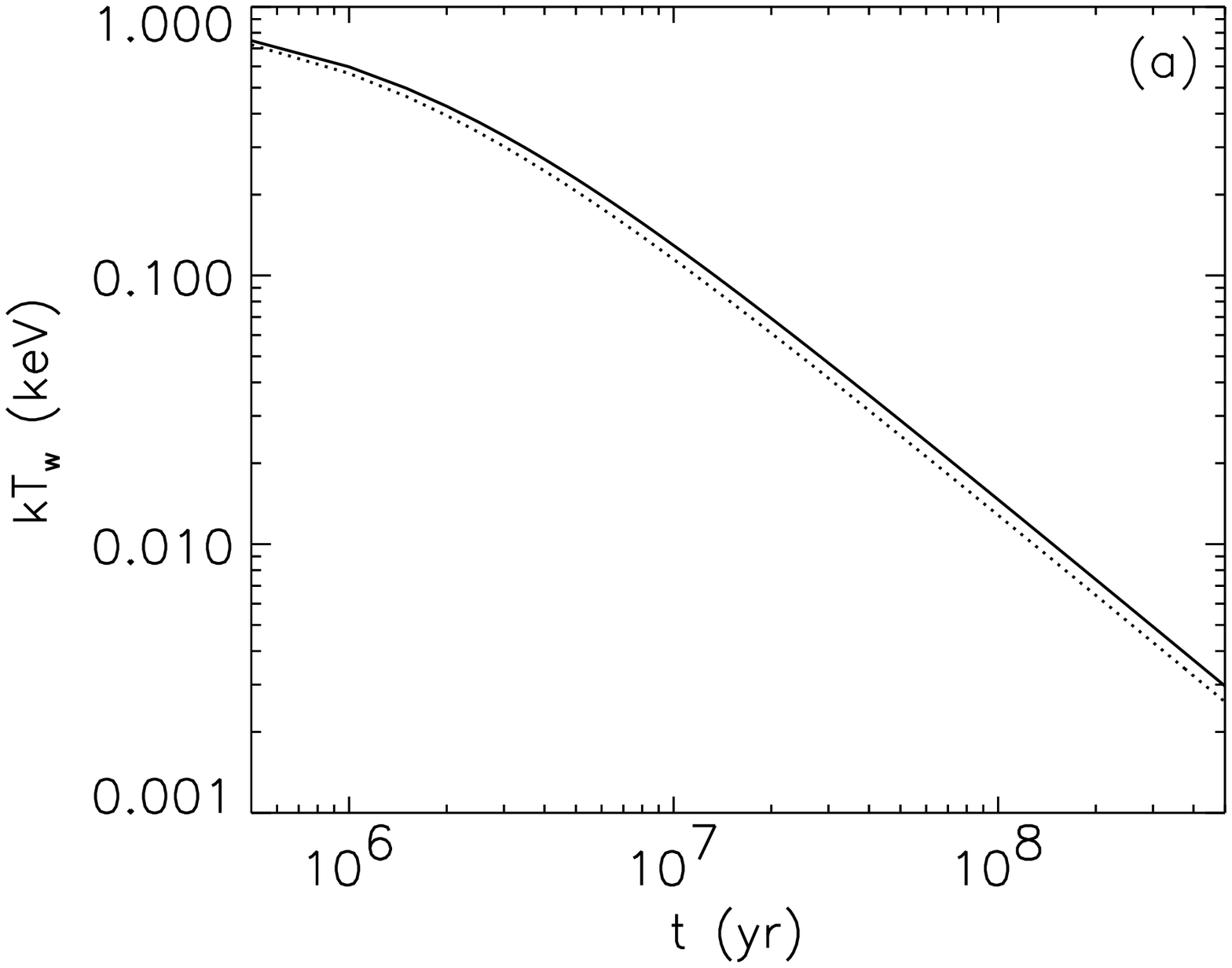} 
 \epsfxsize 0.47\hsize 
 \epsffile{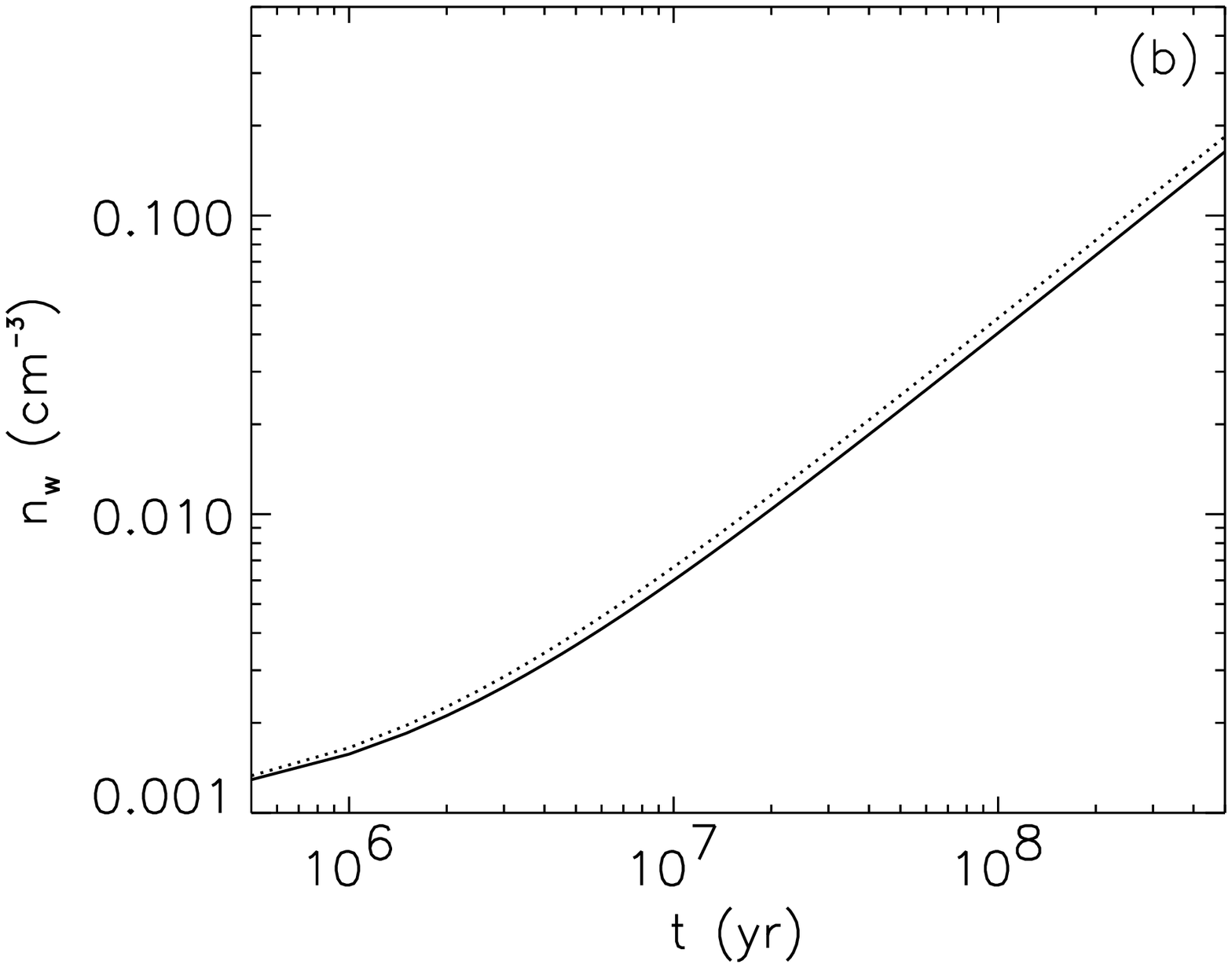}
\end{center}
\begin{center}
 \leavevmode
 \epsfxsize 0.47\hsize 
 \epsffile{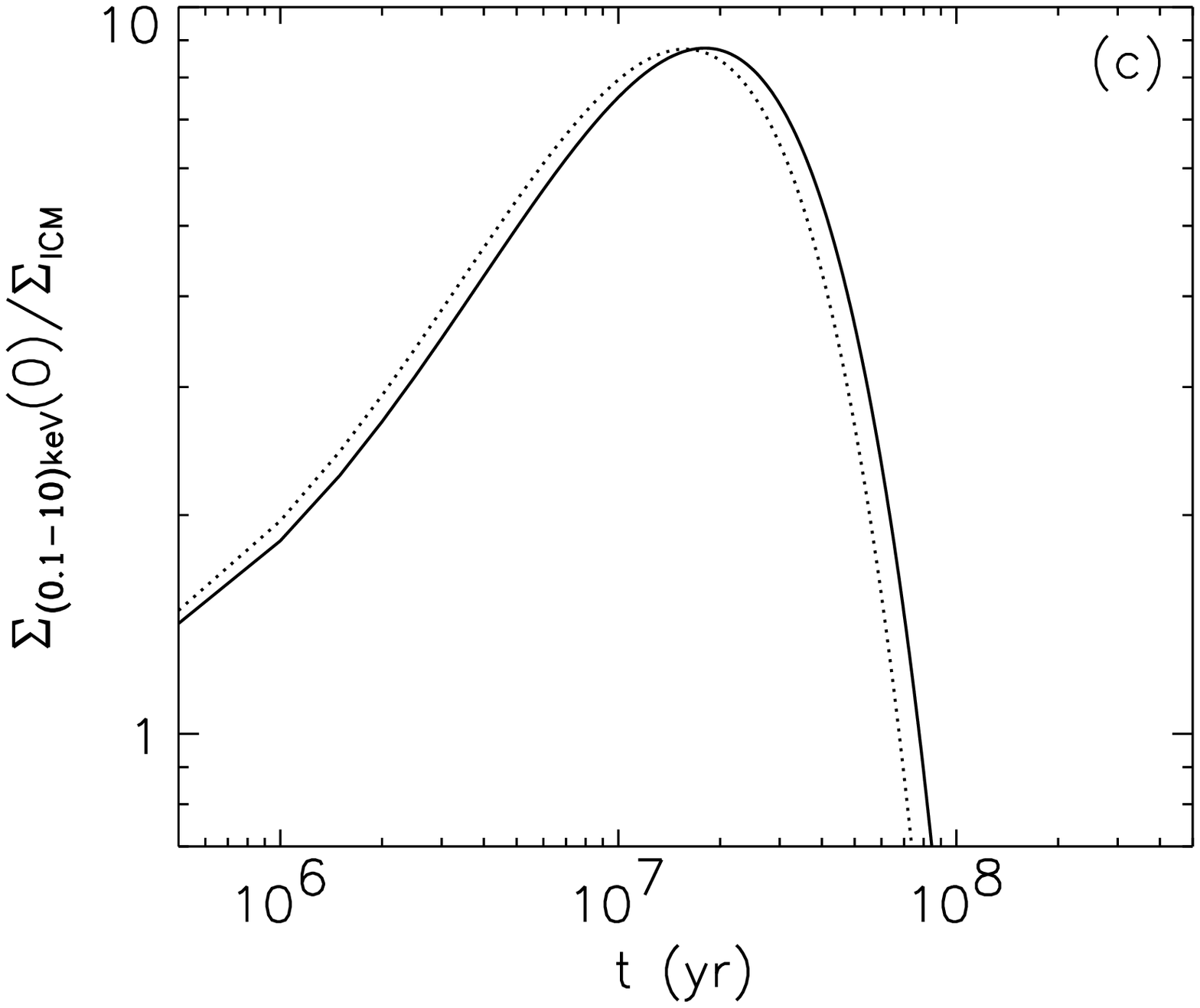}
\end{center}
\caption{Results of the simulation No.~1 for the two extreme impact parameters
($b$) used in the simulations: 10~kpc (solid line) and $R_{\rm acc}$ (dotted
line), which correspond to the closest and furthest parts of the wake to the
galaxy (0.5 and 15~kpc respectively along the accretion axis). The evolution of
the temperature, number density, and central surface brightness in the energy
range (0.1-10)~keV are shown in (a), (b), and (c) respectively.}
\label{first_simul}
\end{figure*}

\subsubsection{The velocity of the galaxy}

To derive the condition of eq.~\ref{mass_con}, an average galaxy velocity was
assumed, which corresponds to the velocity dispersion $\sigma$.  In reality,
the galaxy velocities are distributed around this value, so that in a cluster
there are galaxies which move at lower and larger velocities than $\sigma$. The
accretion radius of fast moving galaxies becomes very small, and they are not
expected to show B-H wakes. It is only the slow moving galaxies which can have
large-scale wakes. Comparing eq.~\ref{sigma_T} and \ref{sound_v}, it can be
concluded that the subsonic velocity regime is the most prolific condition for
B-H wakes.

For the calculations of \S\ref{methodology} the ballistic approximation was
used. The question that arises is whether such an approximation is valid in the
clusters considered.  The ballistic approximation assumes that there is no
interaction between the deflected particles as they stream past the galaxy on
their way to the wake. This assumption is valid only if the kinetic energy of
the deflected particles at any distance $r$ from the galaxy is larger than
their thermal energy. This condition translates to:
\begin{equation} 
v_{\rm gal}^2 +
2 \frac{G M_{\rm gal}}{r} \ga \frac{3}{\gamma} c_{\rm s}^{2}.
\label{conditionI} 
\end{equation} 
Substituting in eq.~\ref{conditionI} the distance $r$ by eq.~\ref{acc_radius}
we find that the ballistic approximation is valid only when:
\begin{equation} v_{\rm gal}^{2} \ga \frac{3 - \gamma}{2 \gamma}
c_{\rm s}^{2} = 0.4 c_{\rm s}^{2} 
\label{conditionII} 
\end{equation}
For any galaxy velocity that obeys eq.~\ref{conditionII}, the
ballistic approximation is correct. Clearly, the lowest limit that
eq.~\ref{conditionII} defines for $v_{\rm gal}$ is always lower than the
average velocity of a cluster galaxy (compare  
equations~\ref{conditionII} and \ref{sigma_T}).

From the above discussion it is clear that the special conditions of the
problem indicate that only slow moving galaxies have wakes with ICM material.
We therefore have to restrain the simulations to subsonic regime, which means
that virialized cluster galaxies should not show the leading bow-shock, which
is a B-H wake's characteristic when the galaxy moves supersonically.  We
decided to use galaxies velocities equal to the local speed of sound and $0.632
c_{s}$, as the condition of eq.~14 requires. The additional advantage of the
subsonic velocities is that no shock waves are formed as stated above, which
would require a different treatment than the one presented here.

\begin{table*} 
\begin{center}
 \caption{B-H accretion : simulations}
 \label{tab:simul}
\end{center}
\begin{tabular}{cccccc}
\hline \hline
No &
$kT_{\rm ICM}$ &
$n_{\rm ICM}$ &
$M_{\rm gal}$ &
$v_{\rm gal}$ &
$R_{\rm acc}$ \\
&
(keV) &
($\times 10^{-3} \ {\rm cm^{-3}}$) &
($\times 10^{12} \ {\rm M_{\sun}}$) &
(${\rm km \ s^{-1}}$) &
(kpc)\\
\hline
1 &
1 &
1 &
2.5 &
330 &
57.3\\

2 &
1 &
1 &
2.5 &
516 &
40.4\\

3 &
2 &
1 &
2.5 &
462 &
28.8\\

4 &
2 &
1 &
2.5 &
730 &
20.2\\

5 &
3 &
1 &
2.5 &
565 &
19.2\\

6 &
3 &
1 &
2.5 &
894 &
13.4\\

\hline
\end{tabular}
\end{table*}

\section{Results}\label{results}

In the following sections we report on the results of the simulation
runs (see Table~\ref{tab:simul}).  Figure~\ref{first_simul} presents
the evolution of the temperature, number density, and central surface
brightness of the wake for the simulation No.~1. The results are shown
for a distance of 0.5~kpc (solid line) and 15~kpc (dotted line) from
the centre of the galaxy along the accretion axis. 

\subsection{The temperature of the wake}

\begin{figure}
 \begin{center}
 \leavevmode
 \epsfxsize 0.99\hsize
 \epsffile{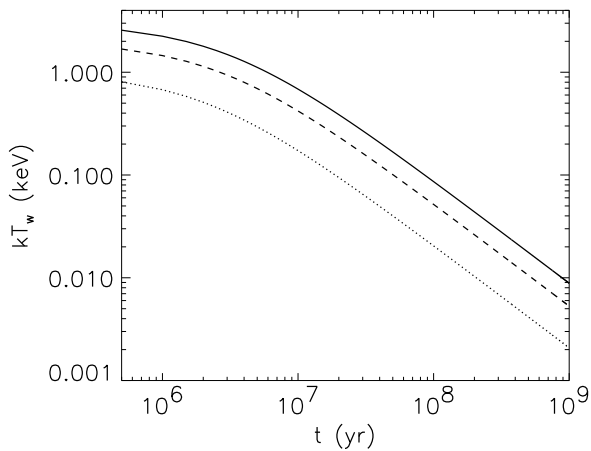}
 \end{center}
\caption{Temperature evolution at a distance of 5~kpc from the
galactic centre in different environments: i) $kT_{\rm ICM}$=1~keV
(dotted line), ii) $kT_{\rm ICM}$=2~keV 
(dashed line), and iii) $kT_{\rm ICM}$=3~keV (solid line). In all
three cases the galaxy
velocity was equal to the local speed of sound.} 
\label{fig:kT}
\end{figure}

As expected, independently of the environment, the trend in temperature is a
continuous cooling with time.  The change of the temperature of the wake at a
distance of 5~kpc from the galaxy in the three different environments studied
is shown in fig.~\ref{fig:kT}.

The signature of the environment is in the rate at which the wake is cooling.
In lower temperature clusters the wake cools down more rapidly than in hotter
environments: in a 3~keV cluster, $kT_{\rm w}$ reaches 10 per cent of the
temperature of the surrounding medium in $3 \times 10^{7}$~yr, while a wake in
a 1~keV cluster needs approximately half the time to reach the same
level. This difference is understood because in low temperature environments,
the external pressure is not as large as in hotter clusters: the pressure
exerted by the accreted particles (eq.~\ref{P_acc}) and $P_{\rm ICM}$ are lower
in cooler clusters. As a result the overdense wake expands more rapidly, and
cools quicker.

\begin{figure}
 \begin{center}
 \leavevmode
 \epsfxsize 0.99\hsize
 \epsffile{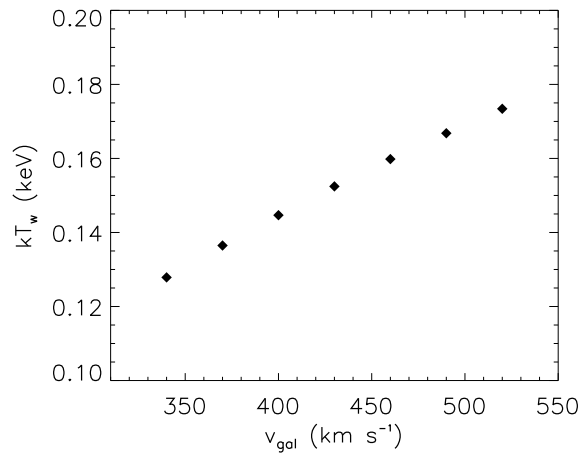}
 \end{center}
\caption{Dependence of the temperature of the wake on the velocity of the
galaxy, in a cluster of $kT_{\rm ICM}$=1~keV, and $n_{\rm ICM}=1 \times 10^{-3} \ {\rm cm^{-3}}$.  The temperature is measured at a distance of 0.5~kpc from the galactic centre, along the accretion axis, and at 
a time of $1 \times 10^{7}$~yr.}
\label{fig:kT_vdep}
\end{figure}

The temperature of the wake was also found to depend on the velocity of the
galaxy. Figure~\ref{fig:kT_vdep} shows that the lower $v_{\rm gal}$, the
quicker the wake cools. This finding can again be understood in terms of the
lower external pressure in cool clusters; the lower $v_{\rm gal}$ is the lower
$P_{\rm acc}$ is.

Along the accretion axis we find a decrease in $kT_{\rm w}$.  At any time, and
any environment the temperature at the extremes of the wake is lower than in
regions closer to the galaxy (by 10-30 per cent; see
fig.~\ref{first_simul}). Such temperature variations might be measurable with
the new x-ray satellites {\it Chandra} and {\it XMM-Newton}.

\subsection{The density of the wake}

\begin{figure}
 \begin{center}
 \leavevmode
 \epsfxsize 0.99\hsize
 \epsffile{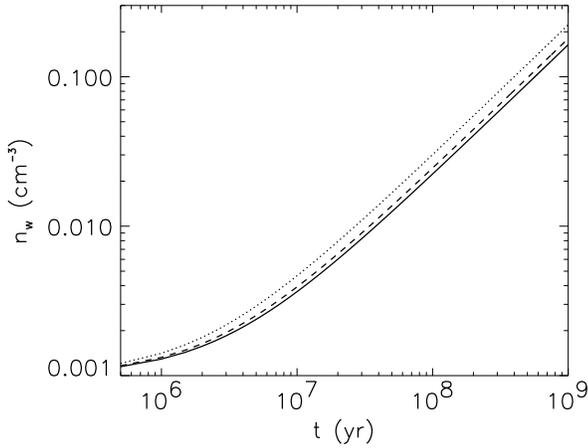}
 \end{center}
\caption{Variations of the number density at a distance of 5~kpc from
the galactic centre along the accretion axis. The results for
different temperature environments are shown: i) $kT_{\rm ICM}$=1~keV
(dotted line), ii) $kT_{\rm ICM}$=2~keV 
(dashed line), and iii) $kT_{\rm ICM}$=3~keV (solid line). The galaxy
velocity was equal to the local sound speed in all three simulations.} 
\label{fig:n}
\end{figure} 

Not surprisingly, the density of the wake increases constantly: in any
environment the increase is approximately one order of magnitude in $\sim
10^{7-8}$~yr. As fig.~\ref{fig:n} shows, in higher temperature clusters the
density of the wake is lower than in cooler ones.  Although the number of ICM
particles accreted into the wake per unit time ($d^{2} z_{\rm acc}$) is larger
in high $kT_{\rm ICM}$ environments (because the galaxy velocity is larger
according to eq.~\ref{z_acc}), the number density of the wake in poorer
environment is larger. This unexpected result can be explained 
because it is the higher flux of particles ($d^{2}z_{\rm acc}/d \sigma$)
that is responsible for the high number density in lower temperature clusters.

\begin{figure}
 \begin{center}
 \leavevmode
 \epsfxsize 0.99\hsize
 \epsffile{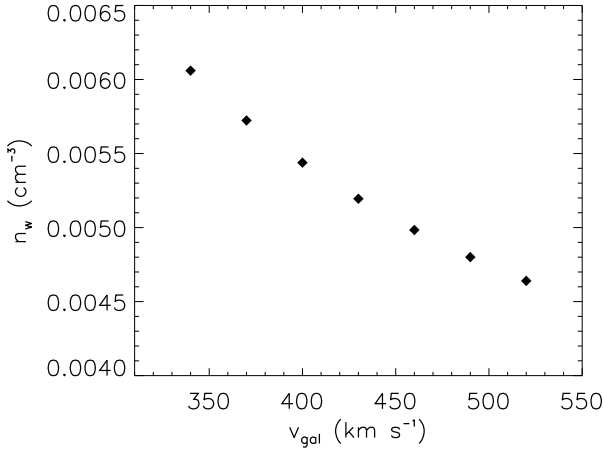}
 \end{center}
\caption{Dependence of the density of the wake on the velocity of the
galaxy, in a cluster of $kT_{\rm ICM}$=1~keV, and $n_{\rm ICM}=1 \times 10^{-3} \ {\rm cm^{-3}}$.  The number density is measured at a distance of 0.5~kpc from the galactic centre, along the accretion axis, and at 
a time of $1 \times 10^{7}$~yr.}
\label{fig:n_vdep}
\end{figure}

Although the effect is not dramatic, in any environment, the wake's
number density has its higher value close to the galaxy, and decreases
with distance from the galaxy along the accretion axis. This result can be
understood in terms of a stronger gravitational potential closer to the
galactic centre. Finally, Figure~\ref{fig:n_vdep} shows the dependence of the
wake's number density on the velocity of the galaxy.

\subsection{Surface brightness}

\begin{figure}
 \begin{center}
 \leavevmode
 \epsfxsize 0.99\hsize
 \epsffile{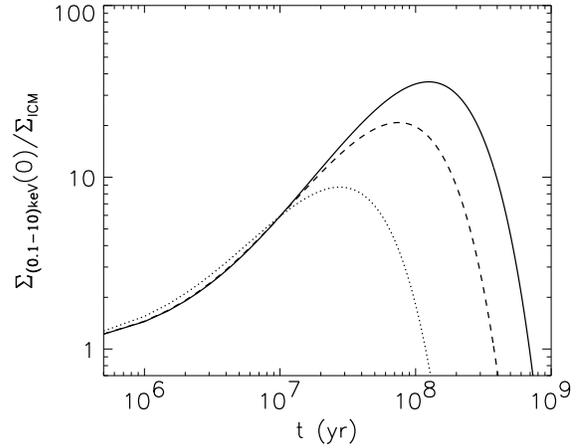}
 \end{center}
\caption{Comparison of the (0.1 - 10)~keV central surface brightness in
different environments: i) $kT_{\rm ICM}$=1~keV (dotted line), ii) $kT_{\rm
ICM}$=2~keV (dashed line), and iii) $kT_{\rm ICM}$=3~keV (solid line). In all
cases the galaxy is moving at the local speed of sound (516, 730, and 894~${\rm
km \ s^{-1}}$ respectively), and the profiles presented in this plot correspond
to a distance of 5~kpc from the core of the galaxy.}
\label{fig:flux}
\end{figure}

Figure~\ref{fig:flux} compares the evolution of the surface brightness
$\Sigma_{(0.1 - 10)~{\rm keV}}(0)$ in a 1~keV, 2~keV, and 3~keV clusters.  The
way that it was calculated is demonstrated in the Appendix~\ref{Lbol}.  The
conversion to any other energy ranges can be performed by applying
eq.~\ref{energy_ranges}.

As fig.~\ref{fig:flux} shows the (0.1-10)~keV central surface brightness in any
environment increases with time until it reaches 10-30 times the brightness of
the surrounding medium. Afterwards, the wake's emission starts declining until
it is immersed in the background, and the wake becomes undetectable. The time
that wakes `disappear' from the x-ray images depends on the richness of the
cluster, with higher temperature clusters being able to retain the wake
signatures for longer time. Additionally, regions which are further away from
the galactic centre fade away quicker than regions closer to the galaxy
(fig.~\ref{first_simul}). The consequence is that a wake becomes shorter with
the course of time. As can be inferred from fig.~\ref{fig:flux}, wakes in
richer clusters live longer.

\subsection{The length of the wake}

\begin{figure}
 \begin{center}
 \leavevmode
 \epsfxsize 0.99\hsize
 \epsffile{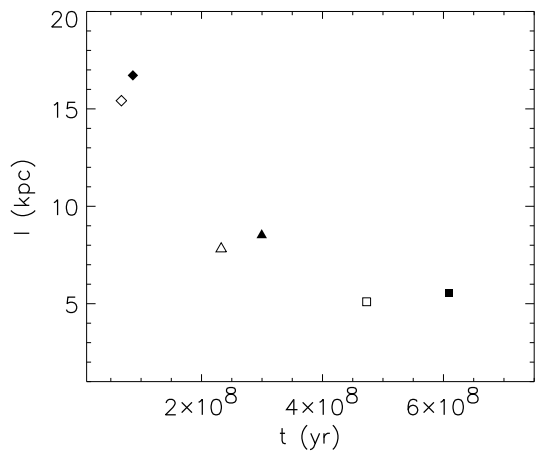}
 \end{center}
\caption{The age of the wake when the surface brightness of the wake's end
becomes equal to the surrounding surface brightness versus the maximum length
in a 1~keV-cluster (diamonds), 2~keV-cluster (triangles), and 3~keV-cluster
(squares). The filled symbols correspond to galaxy velocities equal the local
speed of sound (simulations No 2, 4, and 6 respectively).  The open symbols are
for lower $v_{\rm gal}$ (simulations No 1, 3, and 5).}
\label{fig:length}
\end{figure}

The maximum length that a wake can reach is simply defined by the ballistic
theory as the distance from the galaxy (along the accretion axis) which
corresponds to an impact parameter equal to the accretion radius $R_{\rm acc}$.
However, as it was shown earlier, the surface brightness of the wake reaches
the background level at times that depend on the richness of the environment.
Therefore, it is apparent that the length of the wake depends on the time that
it is observed, and the parameters of the environment which they are in.
Figure~\ref{fig:length} shows for how long the maximum length of a wake is
measurable, before the wake is immersed into the background emission. Although
wakes in richer environments are shorter, they are detectable for longer
periods of time.

\section{Discussion}\label{discussion}

\subsection{Comparison with observations}

It has been demonstrated that the Bondi-Hoyle accretion alone can give rise to
asymmetries in the x-ray images of normal galaxies in relatively rich clusters
of galaxies. However, the scale of these asymmetries is small when compared to
the size of the galaxies: in normal conditions the length of a B-H wake cannot
exceed $\sim$20~kpc.  In a cluster at a redshift of z=0.05 this length
corresponds to $\sim$~12~arcsec, just above the resolution of the {\it ROSAT}
HRI detector. This fact clearly justifies why wakes have been elusive, and why
there are only a few examples reported in the literature.  Even with the new
x-ray satellites {\it Chandra} and {\it XMM-Newton}, wakes will be detectable only in
nearby low temperature clusters.

The best known candidate for a B-H wake might be NGC~1404 in the Fornax cluster
of galaxies. As seen in the {\it ROSAT} PSPC image (Jones et al.  1997), its
wake points away from NGC~1399, which is the central galaxy in the cluster. In
the PSPC data its length appears to be $\sim$25~kpc. From Rangarajan et
al. (1995) we find that at the distance of NGC~1404 the temperature and number
density of the Fornax cluster ICM is $kT_{\rm ICM} \simeq 1.1 \ {\rm keV}$ and
$n_{\rm ICM} \simeq 0.6 \times 10^{-3} \ {\rm cm^{-3}}$. Simulating a galaxy of
$M_{\rm gal} = 2.5 \times 10^{12} \ M_{\sun}$ moving in an environment defined
by the properties of the ICM around NGC~1404, we find that the length of the
wake could be $\sim$17~kpc if the galaxy is moving on the plane of the sky at a
speed of $300 \ {\rm km \ s^{-1}}$, and it cannot exceed 20~kpc if the galaxy
is moving at $500 \ {\rm km \ s^{-1}}$. A first comparison of the PSPC data
with the results of the simulation may suggest that B-H accretion cannot create
a wake as long as the one observed. However, this inconsistency may indicate
that the galaxy is moving at larger speeds than the ones assumed here, owing to
recent infall into the cluster.  Such speculations cannot be verified, and
stringent conclusions cannot be reached before higher spatial and spectral
resolution data of NGC~1404 become available, to accurately constrain the
parameters of the wake from the x-ray observations.

\subsection{Comparison with other simulations}

There have been several studies of the B-H accretion since it was first
explored by Bondi \& Hoyle (1944) in the context of stellar accretion. Hunt
(1971) performed quantitative calculations of subsonic and supersonic accretion
flows for the case of a point source moving in an adiabatic gas.  Sophisticated
3-dimensional simulations have been performed recently (Ruffert 1996 and
references there-in). However, most of these simulations were not designed to
represent the conditions encountered in clusters of galaxies. They explored
either temperature and density regimes which were not representative of
clusters of galaxies, or the galaxy velocities used were too
high. Additionally, the results of these studies cannot be easily converted to
observables, making a comparison with x-ray observations not straightforward.
However, these simulations provide a first picture of the results presented
here. The simulations of a galaxy moving subsonically presented by Ruffert
(1994) show that a downstream overdensity is created. The offset from the
centre of the accretor is small compared to the supersonic
examples. Additionally, as it is clear from the results of Ruffert (1994), in
the supsonic case there is no bow-shock in front of the galaxy. On the other
hand the bow-shock is the most prominent structure generated by the B-H
accretion onto a supersonically moving body. As stated by Ruffert (1994), a B-H
wake in the subconic case {\it ``is the result of the superposition of the slow
radial flow into the accretor and the relative motion between the accretor and
the medium''}.

Recently, more in depth predictions of the ICM/ISM interactions in clusters of
galaxies have been presented by Stevens et al.~(1999). Their simulations allow
both dynamical processes to take place: ICM is deflected and concentrated
behind the moving galaxy, and ram pressure strips the galactic ISM. The result
is the creation of downstream density enhancements, which consist of both
galactic and intracluster material. Unfortunately, this study does not assess
the fraction of each component expected in the wake for a variety of
environmental conditions.  As a result, one cannot confidently estimate whether
the wakes we observe consist of galactic or ICM material.

However, these simulations give the first indication that B-H accretion might
dominate in low temperature clusters, in agreement with our results, and ram
pressure creates wakes in richer clusters. If one was to compare the results of
the present study with the simulations of Stevens et al., he would have chosen
to use their simulation No.~1b, for the sake of consistency. Indeed, simulation
No.~1b presents the results of a galaxy/ICM interaction in a cluster with
temperature $kT_{\rm ICM}$=1~keV, and density $n_{\rm ICM} \sim 4 \times
10^{-4} \ {\rm cm^{-3}}$. These parameters of the ICM correspond well to the
conditions studied here. Generally, there is a good agreement between both
studies: they both predict an increase of approximately two orders of magnitude
in the $n_{\rm w}$, with a simultaneous temperature drop by two orders of
magnitude at 5~kpc along the accretion axis and after $5 \times 10^{8}$~yr (see
fig.~3 in Stevens et al.). The agreement between the two studies may lead to
the false conclusion that both simulations describe the same physical process
successfully. However, special care must be paid, because the galaxy velocity in
Stevens et al. is supersonic with $v_{\rm gal}=960 \ {\rm km \ s^{-1}}$. As it
was proven in the previous sections, such velocity regimes cannot produce
large-scale B-H wakes, because the accretion radius is too small. However, the
lack of striking differences make us conclude that B-H dominates in No.~1b
simulation of Stevens et al.

\section{Summary and Conclusions}

The motion of a body through a gaseous medium results in the creation of an
overdense, cool wake: the gravitational attraction of the body deflects the
surrounding medium's particles, which end up being concentrated behind it.
This process (Bondi-Hoyle accretion) was first explored by Bondi \& Hoyle
(1944) in the context of stellar accretion. The aim of this paper was to
investigate if this process is at work in clusters of galaxies, which clusters
are the best hosts of B-H wakes, and if such features can be seen in the x-ray
data of clusters of galaxies.

The main results of our calculations can be summarized as follows:

\begin{itemize}

 \item{In clusters of galaxies, the ICM can be concentrated behind the
 subsonically moving galaxies, although the results of B-H accretion are
 expected to be more dramatic when the galaxies move supersonically.}

 \item{Large-scale B-H wakes can be found only in low temperature clusters,
 behind slow moving, and massive galaxies.}

 \item{A stable situation is not reached within reasonable timescales: the wake
 is created, its temperature decreases and its density increases
 constantly.}

 \item{The smaller the temperature of the ICM, the longer the wake
 is. However, wakes in richer clusters live longer and are brighter. Such
 features will be easily detectable with the {\it Chandra} and {\it XMM-Newton}
 satellites in nearby, poor clusters of galaxies.} 

 \item{The central, x-ray surface brightness of the wake can reach 10-30 times
 the brightness of the surrounding medium.}

 \item{The properties of a B-H wake depend on the velocity of the galaxy. Fast
 moving galaxies, for example, have hotter wakes. On the other hand, galaxies
 with pronounced B-H wakes should not have bow shocks, because their motion
 is subsonic.}

\end{itemize}

\section*{ACKNOWLEDGMENTS}

We would like to thank M.~Cropper for useful discussion, and J.~Lee for
critical reading of the manuscript. Special thanks to C.~Lee for the
encouragement to undertake this investigation.

\appendix

\section{Bolometric luminosity and Surface brightness distribution}\label{Lbol}

The bolometric luminosity $L_{\rm bol}(x,t)$ used in eq.~\ref{energy}
to describe the energy radiated away, is found by integrating the
bremsstrahlung emissivity over the volume of the wake ($V_{\rm w}$) at
any time $t$:
\begin{eqnarray}
L_{\rm bol}(x,t)= \nonumber \\
6.8 \times 10^{-38} \ \frac{k^{1/2}}{h} \ \overline{g}_{\rm
B} \int kT_{\rm w}^{1/2}(x,t)  n_{\rm w}(x,t)^{2} dV_{\rm w}
\label{brems}
\end{eqnarray}

If we assume that the emitting region is a cylinder, the gas in the
wake is uniformally distributed, and it emits isotropically, the
surface brightness [$\Sigma_{\rm E_1 - E_2}(\theta)$] at any distance $\theta$ from the
accretion axis is the integrated bremsstrahlung
emissivity along a column at the projected distance $\theta$ from the accretion
axis:
\begin{eqnarray}
\Sigma_{\rm E_1 - E_2}(\theta)= \nonumber \\ 
6.8 \times 10^{-38} n_{\rm w}^{2} T_{\rm
w}^{-1/2} \overline{g}_{\rm B} \ I_{\rm E_1 - E_2} \ \int_{r= \theta}^{r=R_{\rm
w}(x,t)} \frac{rdr}{\sqrt{r^{2} - \theta^{2}}}
\label{surf_bri}
\end{eqnarray} 
where $I_{\rm E_1 - E_2}$ the integral which defines the energy range in which
$\Sigma_{\rm E_1 - E_2}(\theta)$ is calculated, and it is given by:
\begin{equation}
I_{\rm E_1 - E_2}=\int_{\nu_{1}}^{\nu_{\rm 2}} \exp{ \left(- \frac{h
\nu}{kT_{\rm w}} \right)} d \nu
\label{energy_ranges}
\end{equation}

Integrating eq.~\ref{surf_bri} we find that the surface brightness at
any point $\theta$ is:
\begin{equation}
\Sigma_{\rm E_1 - E_2}(\theta) = \Sigma_{\rm E_1 -E_2}(0) \left[1 - \left( \frac{\theta}{R_{\rm
w}(x,t)} \right)^{2} \right]^{1/2}
\label{surf_bri_distr}
\end{equation}
where the central surface brightness $\Sigma_{\rm E_1 - E_2}(0)$ is given by:
\begin{equation}
\Sigma_{\rm E_1 - E_2}(0)=13.6 \times 10^{-38} n_{\rm w}^{2} T_{\rm w}^{-1/2}
\overline{g}_{\rm B} \ I_{\rm E_1 - E_2} \ R_{\rm w}(x,t)
\label{central_surf_bri}
\end{equation}


\begin{thebibliography}{}


\bibitem[\protect\citename{Bondi} 1952]{} Bondi H., 1952,
  MNRAS, 112, 195 
\bibitem[\protect\citename{Bondi, Hoyle} 1996]{} Bondi H., Hoyle F., 1944,
  MNRAS, 104, 273 
\bibitem[\protect\citename{Hunt} 1971]{} Hunt R., 1971,
  MNRAS, 154, 141 
\bibitem[\protect\citename{Jones} 1997]{} Jones C., Stern C., Forman W., Breen
J., David L., Tucker W., Franx M., 1997, ApJ, 482, 143
\bibitem[\protect\citename{Rangarajan} 1995]{} Rangarajan F.V.N., White D.A.,
Ebeling H., Fabian A.C., 1995, MNRAS, 277, 1047
\bibitem[\protect\citename{Ruffert } 1994]{} Ruffert M., A\&AS, 106, 505
\bibitem[\protect\citename{Ruffert } 1996]{} Ruffert M.,
  1996, A\&A, 311, 817 
\bibitem[\protect\citename{Stevens et al.} 1999]{} Stevens I.R., Acreman D.M.,
Ponman T.J., 1999, MNRAS, 310, 663
\bibitem[\protect\citename{White et al.} 1997]{} White
  D. A., Jones C., Forman W., 1997, MNRAS, 292,419
  \bibitem[\protect\citename{Wu et al.} 1998]{} Wu X.-P., Fang L.-Z., Xu W.,
  1998, A\&A, 338, 813

\end{thebibliography}
\end{document}